\documentclass[amsmath,amssymb,twocolumn,prc,showpacs,nofootinbib,floatfix,superscriptaddress]{revtex4-1}
\usepackage{graphics}
\usepackage{dcolumn}
\usepackage{bm}
\usepackage{longtable}
\usepackage[dvips]{graphicx}
\usepackage{dcolumn}
\usepackage{footnote}
\usepackage{amssymb}
\usepackage{multirow}
\usepackage[hidelinks]{hyperref}
\usepackage{xcolor}
\hypersetup{
    colorlinks,
    linkcolor={red!50!black},
    citecolor={blue!50!black},
    urlcolor={blue!80!black}
}
\begin{document}

\title{Revalidation of the isobaric multiplet mass equation for the $A=20$ quintet}
\author{B. E. Glassman}
\email{glassman@nscl.msu.edu}
\affiliation{Department of Physics and Astronomy, Michigan State University, East Lansing, Michigan 48824, USA}
\affiliation{National Superconducting Cyclotron Laboratory, Michigan State University, East Lansing, Michigan 48824, USA}
\author{D. P\'{e}rez-Loureiro}
\email{perezlou@nscl.msu.edu}
\affiliation{National Superconducting Cyclotron Laboratory, Michigan State University, East Lansing, Michigan 48824, USA}
\author{C. Wrede}
\email{wrede@nscl.msu.edu}
\affiliation{Department of Physics and Astronomy, Michigan State University, East Lansing, Michigan 48824, USA}
\affiliation{National Superconducting Cyclotron Laboratory, Michigan State University, East Lansing, Michigan 48824, USA}
\author{J. Allen}
\affiliation{Department of Physics, University of Notre Dame, Notre Dame, Indiana 46556, USA}
\author{D. W. Bardayan}
\affiliation{Department of Physics, University of Notre Dame, Notre Dame, Indiana 46556, USA}
\author{M. B. Bennett}
\affiliation{Department of Physics and Astronomy, Michigan State University, East Lansing, Michigan 48824, USA}
\affiliation{National Superconducting Cyclotron Laboratory, Michigan State University, East Lansing, Michigan 48824, USA}
\author{B.~A.~Brown}
\affiliation{Department of Physics and Astronomy, Michigan State University, East Lansing, Michigan 48824, USA}
\affiliation{National Superconducting Cyclotron Laboratory, Michigan State University, East Lansing, Michigan 48824, USA}
\author{K. A. Chipps}
\affiliation{Oak Ridge National Laboratory, Oak Ridge, Tennessee 37831, USA}
\affiliation{Department of Physics and Astronomy, University of Tennesssee, Knoxville, Tennessee 37996, USA}
\author{M. Febbraro}
\affiliation{Oak Ridge National Laboratory, Oak Ridge, TN 37831, USA}
\affiliation{Department of Physics and Astronomy, University of Tennesssee, Knoxville, Tennessee 37996, USA}
\author{C. Fry}
\affiliation{Department of Physics and Astronomy, Michigan State University, East Lansing, Michigan 48824, USA}
\affiliation{National Superconducting Cyclotron Laboratory, Michigan State University, East Lansing, Michigan 48824, USA}
\author{M.~R. Hall}
\affiliation{Department of Physics, University of Notre Dame, Notre Dame, Indiana 46556, USA}
\author{O. Hall}
\affiliation{Department of Physics, University of Notre Dame, Notre Dame, Indiana 46556, USA}
\author{S. N. Liddick}
\affiliation{Department of Chemistry, Michigan State University, East Lansing, Michigan 48824, USA}
\affiliation{National Superconducting Cyclotron Laboratory, Michigan State University, East Lansing, Michigan 48824, USA}
\author{P.~O'Malley}
\affiliation{Department of Physics, University of Notre Dame, Notre Dame, Indiana 46556, USA}
\author{W. Ong}
\affiliation{Department of Physics and Astronomy, Michigan State University, East Lansing, Michigan 48824, USA}
\affiliation{National Superconducting Cyclotron Laboratory, Michigan State University, East Lansing, Michigan 48824, USA}
\author{S. D. Pain}
\affiliation{Oak Ridge National Laboratory, Oak Ridge, Tennessee 37831, USA}
%
%\author{C. Prokop}
%\affiliation{Department of Chemistry, Michigan State University, East Lansing, Michigan 48824, USA}
%\affiliation{National Superconducting Cyclotron Laboratory, Michigan State University, East Lansing, Michigan 48824, USA}
%
\author{S. B. Schwartz}
\affiliation{Department of Physics and Astronomy, Michigan State University, East Lansing, Michigan 48824, USA}
\affiliation{National Superconducting Cyclotron Laboratory, Michigan State University, East Lansing, Michigan 48824, USA}
\author{P. Shidling}
\affiliation{Cyclotron Institute, Texas A \& M University College Station, Texas 77843, USA}
\author{H. Sims}
\affiliation{University of Surrey, GU2 7XH, Guildford, UK}
\author{P. Thompson}
\affiliation{Oak Ridge National Laboratory, Oak Ridge, Tennessee 37831, USA}
\affiliation{Department of Physics and Astronomy, University of Tennesssee, Knoxville, Tennessee 37996, USA}
\author{H. Zhang}
\affiliation{Department of Physics and Astronomy, Michigan State University, East Lansing, Michigan 48824, USA}
\affiliation{National Superconducting Cyclotron Laboratory, Michigan State University, East Lansing, Michigan 48824, USA}
% Others: Schatz group (used vacuum system); Buhro few hours, Naviliat and post-doc hung around; Liddick group?
%
\date{\today}

\begin{abstract}

An unexpected breakdown of the isobaric multiplet mass equation in the $A=20$, $T=2$ quintet was recently reported, presenting a challenge to modern theories of nuclear structure. In the present work, the excitation energy of the lowest $T = 2$ state in $^{20}$Na has been measured to be $6498.4 \pm 0.2_{\textrm{stat}} \pm 0.4_{\textrm{syst}}$ keV by using the superallowed $0^+ \rightarrow 0^+$ beta decay of $^{20}$Mg to access it and an array of high-purity germanium detectors to detect its $\gamma$-ray deexcitation. This value differs by 27 keV (1.9 standard deviations) from the recommended value of $6525 \pm 14$ keV and is a factor of 28 more precise. The isobaric multiplet mass equation is shown to be revalidated when the new value is adopted.

\end{abstract}


\maketitle

Isospin symmetry considers the proton and the neutron to be degenerate states of the same particle motivated by their similar masses and similar interactions via the strong nuclear force \cite{he32zp,wi37pr}. In reality, isospin symmetry is broken by the different charges and masses of the two particles. These evident differences can be accounted for by using first-order perturbation theory, restoring the broad utility of isospin symmetry in nuclear structure and nuclear astrophysics \cite{ma14npa}. In particular, the nuclear states that are members of a multiplet of isospin $T$ are not perfectly degenerate, but their mass excesses $\Delta$ can be related by the simple Isobaric Multiplet Mass Equation (IMME) \cite{wi57,we58pr},

\begin{equation}
\Delta(T_z) = a + bT_z + cT_z^2.
\label{eq: IMME}
\end{equation}

\noindent In Eq. (\ref{eq: IMME}), $T_z = (N-Z)/2$ is the projection of the isospin and $a$, $b$, and $c$ are coefficients that can be calculated theoretically, or determined empirically by using the IMME to fit the experimentally determined mass excesses of the multiplet \cite{ma14npa}. A poor fit indicates a breakdown of the IMME, which can be quantified by nonzero $d$ and $e$ coefficients to cubic or quartic terms in $T_z$, respectively. Charge-dependent nuclear forces, second-order Coulomb effects, and three-body interactions have been predicted to produce $d$ coefficients with magnitudes lower than $\approx 1$ keV \cite{he69pr,ay81prc,ja69npa,be70plb,au72npa} provided the most neutron-deficient member of the multiplet is bound and mixing between states of different isospin is weak; values beyond this represent a significant and unexpected breakdown.

The $A=20$, $T=2$ multiplet consisting of the lowest-energy $T=2$ states in $^{20}$Mg, $^{20}$Na, $^{20}$Ne, $^{20}$F, and $^{20}$O is the lightest quintet for which the most neutron-deficient ($T_z = -2$) member ($^{20}$Mg) is bound \cite{ma14npa} and, in addition, isospin mixing is expected to be weak in this system \cite{ga14prl}. Independence from these potentially dominant effects makes the $A=20$ quintet an ideal testing ground for more subtle deviations from the IMME \cite{ga64prl,pe65pl,do65prl,ro74prl,mi76prc,tr76plb,mo79prl,an82jpg,sh99epj,ga07prc,wr10prc,fo12prc,ga14prl}. Recently, the ground state mass of $^{20}$Mg was measured to high precision, enabling the most stringent test of the IMME in the $A=20$, $T=2$ multiplet \cite{ga14prl} so far. The authors concluded that the IMME is violated, presenting a major unexpected challenge to modern shell-model calculations. However, the other masses and excitation energies in the multiplet were necessarily adopted from evaluations of existing data; inaccurate adopted values could potentially lead to erroneous conclusions about the validity of the IMME. Therefore, it seems prudent to check, and improve upon, the other values.

\begin{figure}
\includegraphics[width=0.5\textwidth]{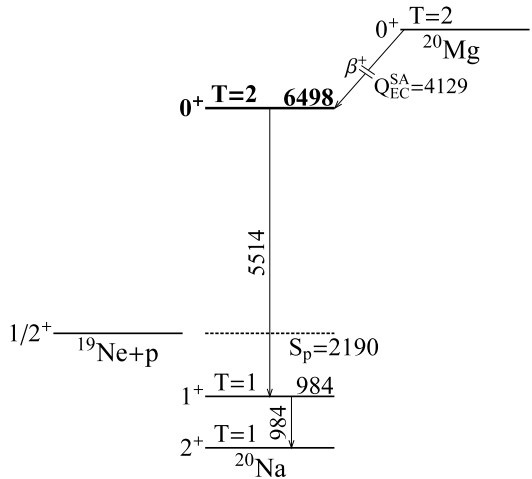}
\caption{Simplified $^{20}$Mg $\beta$-decay scheme focusing on the transitions relevant to the present work. Energies are shown in units of keV.}
\label{fig: scheme}
\end{figure}

The largest uncertainty, by far, is the 14 keV uncertainty associated with the mass excess of the lowest $T = 2$ state in $^{20}$Na \cite{ma14npa}, which is based on measurements of the energies of $^{20}$Mg $\beta$-delayed protons \cite{mo79prl,go92prc,pi95npa}. Although $T =1/2$ proton emission from a $T = 2$ state to produce $T = 1/2$ $^{19}$Ne is forbidden by conservation of isospin, the $^{20}$Na state is sufficiently proton unbound (by $>4$ MeV) that the proton emission proceeds anyways and is, in fact, the dominant decay mode. Nevertheless, isospin suppression of the proton width should be strong enough to provide an observable $\gamma$-decay branch of a few percent. If the $\gamma$ rays from the lowest $T = 2$ state of $^{20}$Na could be observed by using high-resolution $\gamma$-ray detectors then the excitation energy could be determined to much higher precision and accuracy. Adding the excitation energy to the recently determined precise ground-state mass of $^{20}$Na \cite{wr10prc,wa12cpc} would provide the mass of the lowest $T = 2$ state, which could then be used for an improved IMME test. We measured the excitation energy of the lowest $T = 2$ state in $^{20}$Na by using the $\beta$-delayed $\gamma$ decay of $^{20}$Mg (Fig. \ref{fig: scheme}), which has only been measured once before yielding a single $^{20}$Na $\gamma$-ray transition from a low-lying bound state \cite{pi95npa}.

The experiment was carried out at Michigan State University's National Superconducting Cyclotron Laboratory (NSCL), which provided a fast radioactive $^{20}$Mg beam by using projectile fragmentation of a 170 MeV/u, 60 pnA $^{24}$Mg primary beam from the Coupled Cyclotron Facility, incident upon a 961 mg/cm$^2$ $^{9}$Be transmission target. The $^{20}$Mg ions were separated from other fragmentation products by magnetic rigidity by using the A1900 fragment separator, which incorporated a 594 mg/cm$^2$ Al wedge \cite{mo03nim}. Rates of up to 4000 $^{20}$Mg ions s$^{-1}$ were delivered to the experimental setup. Beam ions were cleanly identified by combining the time of flight from a scintillator at the focal plane of the A1900 to a 300-$\mu$m-thick silicon detector located $\approx 70$ cm upstream of the counting station with the energy loss in the latter. Between runs, the beam intensity was attenuated and the composition was sampled to avoid excessive radiation damage to the Si detector, which was extracted while running. The average composition of the beam delivered to the experiment was found to be 43 \% $^{20}$Mg with the contaminant isotones $^{18}$Ne (28 \%), $^{17}$F (7 \%), $^{16}$O (19 \%), and $^{15}$N (3 \%). The $^{20}$Mg ions were implanted in a 25-mm thick plastic scintillator. The scintillator recorded the ion implantations and their subsequent $\beta$ decays. The Segmented Germanium Array (SeGA) of high-purity Ge detectors \cite{mu01nim} surrounded the scintillator in two coaxial 13-cm radius rings consisting of 8 germanium detectors apiece and it was used to detect $\gamma$ rays. The NSCL digital data acquisition was employed \cite{pr14nim}.

The SeGA spectra were gain matched to produce cumulative spectra by using the strong $\gamma$-ray lines from room-background activity with transition energies of $1460.851 \pm 0.006$ keV (from $^{40}$K decay) \cite{ca04nds} and $2614.511 \pm 0.010$ keV (from $^{208}$Tl decay) \cite{ma07nds} as reference points, providing an \emph{in situ} first-order energy calibration. In order to reduce the room-background contribution to the $\gamma$-ray spectra, a $\beta$-coincident $\gamma$-ray spectrum was produced by requiring coincidences with $\beta$ particle signals from the implantation scintillator. Lines with well known transitions energies of $1633.602 \pm 0.015$, $3332.84 \pm 0.20$, $6129.89 \pm 0.04$, $8239 \pm 4$, and $8640 \pm 3$ keV \cite{ti98npa,ti93npa} from the $\beta$-delayed $\gamma$ (and $\alpha$-$\gamma$) decays of $^{20}$Na (the daughter of $^{20}$Mg $\beta$-decay) were observed with high statistics and used together with the two room-background lines for a more extensive energy calibration. Corrections for the energy carried by daughter nuclei recoiling from $\gamma$-ray emission were applied throughout the calibration procedure.

%\begin{figure}
%\includegraphics[width=0.5\textwidth]{full_spectrum}
%\caption{(Color online) $^{20}$Mg \$beta$-delayed \$gamma$ ray spectrum.}
%\label{fig: spectrum}
%\end{figure}

\begin{figure}
\includegraphics[width=0.5\textwidth]{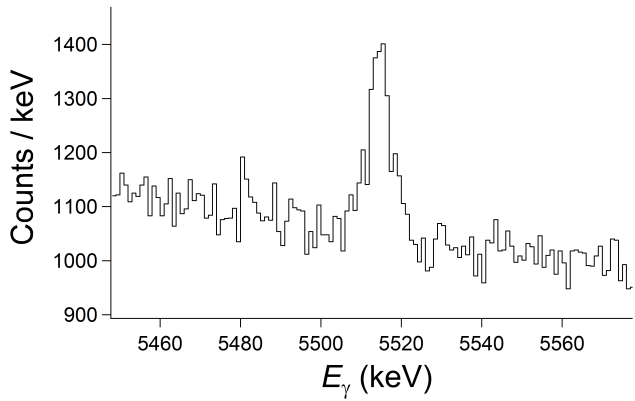}
\caption{$\beta$-coincident $\gamma$-ray spectrum focusing on the 5514 keV peak.}
\label{fig: singles_peak}
\end{figure}

\begin{figure}
\includegraphics[width=0.5\textwidth]{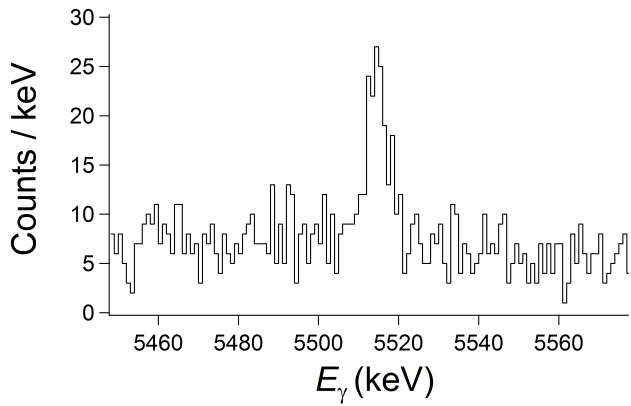}
\caption{$\beta$-coincident $\gamma$-ray spectrum, with additional coincidence gating condition on the 984 keV $^{20}$Na $\gamma$-ray peak (Fig. \ref{fig: 984_peak}), focusing on the 5514 keV peak.}
\label{fig: coinc_peak}
\end{figure}

\begin{figure}
\includegraphics[width=0.5\textwidth]{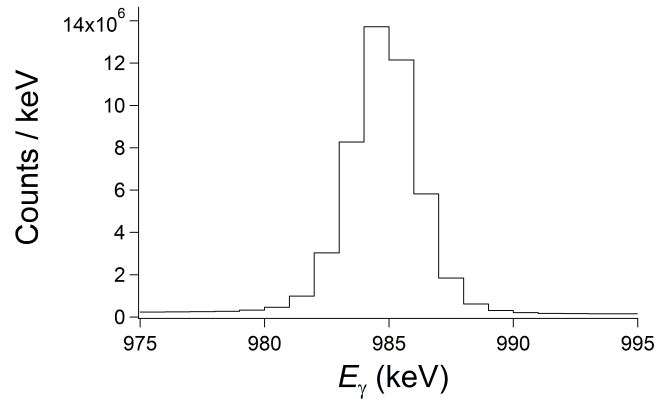}
\caption{$\beta$-coincident $\gamma$-ray spectrum focusing on the 984 keV peak.}
\label{fig: 984_peak}
\end{figure}

Clear evidence for a new $\gamma$ ray at a laboratory energy of $5513.9 \pm 0.2_\textrm{stat} \pm 0.4_\textrm{syst}$ keV was observed (Fig. \ref{fig: singles_peak}). This peak was confirmed to be from a high-lying level of $^{20}$Na by placing a coincidence condition on the well-known 984 keV $^{20}$Na $\gamma$-ray transition (Fig. \ref{fig: coinc_peak}) \cite{pi95npa}, showing that the 5514 keV $\gamma$ ray feeds the 984 keV level. The latter peak was observed at a laboratory energy of $983.70 \pm 0.00_\textrm{stat} \pm 0.10_\textrm{syst}$ keV (Fig. \ref{fig: 984_peak}). The statistical uncertainties associated with the energies of these peaks were determined by fitting them with Gaussian and exponentially modified Gaussian functions and a linear background. The systematic uncertainty was dominated by uncertainties associated with the energy calibration including the adopted nuclear data and the peak-fitting procedure, which was varied to test the sensitivity to details. Applying the recoil correction yields values of $983.73 \pm 0.00_\textrm{stat} \pm 0.10_\textrm{syst}$ keV and $5514.7 \pm 0.2_\textrm{stat} \pm 0.4_\textrm{syst}$ keV for the transition energies.

Adding the 984 and 5515 keV $\gamma$-ray transition energies yields a $^{20}$Na excitation energy of $6498.4 \pm 0.2_\textrm{stat} \pm 0.4_\textrm{syst}$ keV for the observed state (for subsequent calculations we combine the two uncertainties in quadrature and use the value $6498.4 \pm 0.5$ keV). There are two pieces of evidence that this state corresponds to the lowest $T = 2$ state of $^{20}$Na. First, it would be surprising to observe $\beta$-delayed $\gamma$ decays from a $T = 1$ $^{20}$Na state that is unbound by several MeV, but such an observation is not unexpected for a $T = 2$ state in a $T_z = -1$ nuclide because, as discussed above, proton emission is isospin forbidden. Second, the lowest $T = 2$ state is predicted by the $sd$ shell model to have a dominant decay branch to the 984 keV state, as we observed (the other primary branches are expected to be more than an order of magnitude weaker and were not observed).

The present excitation energy of the lowest $T = 2$ state is 27 keV (1.9 standard deviations) lower than the value of $6525 \pm 14$ keV from the most recent data evaluation \cite{ma14npa}, which was based on several measurements of $^{20}$Mg $\beta$-delayed proton emission \cite{mo79prl,go92prc,pi95npa}. The present value is also 28 times more precise. Adopting our new value for the excitation energy of the lowest $T = 2$ state in $^{20}$Na and the recently measured \cite{wr10prc} and evaluated \cite{wa12cpc} $^{20}$Na ground-state mass excess of $6850.6 \pm 1.1$ keV yields a mass excess of $13349.0 \pm 1.2$ keV for the $T = 2$ state, where the uncertainties have been combined in quadrature.

\begin{figure}
\includegraphics[width=0.5\textwidth]{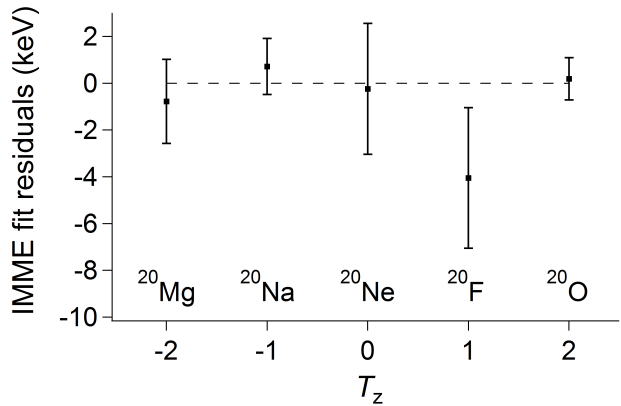}
\caption{Residuals for the quadratic IMME fit (Equation \ref{eq: IMME}) of the $A = 20$, $T = 2$ quintet from the present work (Tables \ref{tab:IMMEinput} and \ref{tab:IMMEoutput}).}
\label{fig: IMME_residuals}
\end{figure}

We have adopted this value together with the recommended values for $^{20}$O, $^{20}$F, and $^{20}$Ne from Ref. \cite{ma14npa}, and the recently measured precise value of the $^{20}$Mg mass from Ref. \cite{ga14prl} to test the IMME in the $A=20$, $T=2$ multiplet (Table \ref{tab:IMMEinput}). In addition to applying a standard quadratic IMME fit [Eq.(\ref{eq: IMME})], we fit the data using a cubic fit, a quartic fit, and a quartic fit with the cubic coefficient set to zero in order to gauge the potential need for extra terms. The coefficients derived from the fits are reported together with the goodness of the fits in Table \ref{tab:IMMEoutput}. The quadratic IMME is found to provide an excellent fit to the data, yielding $\chi^2/\nu = 2.4/2$. The small residuals of the fit (Fig. \ref{fig: IMME_residuals}) reflect the improvement in the precision and accuracy of the $T_z = -1$ member of the multiplet, $^{20}$Na, which is now one of the two most precisely known members of the quintet. When a cubic term is added, the $d$ coefficient is found to be $0.8 \pm 0.5$ keV, which is less than 1 keV, consistent with zero, and consistent with the value of $-0.1$ keV predicted by the shell model \cite{ga14prl} within two standard deviations. In contrast to Ref. \cite{ga14prl}, which reported $d = 2.8 \pm 1.1$ keV leading to the assertion that the IMME is violated, we find that the IMME is revalidated. Therefore, there is no need to introduce exotic new subatomic theories to explain the current experimental data.

\begin{table}
\caption{\label{tab:IMMEinput} IMME input mass excesses, $\Delta_{T=2}$, for the lowest $A=20$, $T=2$ quintet, including the constituent ground-state mass excesses $\Delta_{\textrm{g.s.}}$ and excitation energies $E_x$. The values for the $T_z = +2, +1, 0$ states and the value of $\Delta_{\textrm{g.s.}}$ for the $T_z = -1$ state are from Ref. \cite{ma14npa}. The value for the $T_z = -2$ state is from Ref. \cite{ga14prl}. The value of $E_x$ for the $T_z = -1$ state is from the present work.}
\begin{ruledtabular}
\begin{tabular}{c c c c c}
Nuclide  & $T_z$ & $\Delta_{\textrm{g.s.}}$ (keV) & $E_x$ (keV) & $\Delta_{T=2}$ (keV) \\
\hline
$^{20}$O      & +2   & 3796.2(9)        &             & 3796.2(9)   \\
$^{20}$F      & +1   & $-17.463(30)$    & 6521(3)     & 6503(3)     \\
$^{20}$Ne     &  0   & $-7041.9306(16)$ & 16732.8(28) & 9690.9(28)  \\
$^{20}$Na     & $-1$ & 6850.6(11)       & 6498.4(5)   & 13349.0(12) \\
$^{20}$Mg     & $-2$ & 17477.7(18)      &             & 17477.7(18) \\
\end{tabular}
\end{ruledtabular}
\end{table}

\begin{table}
\caption{\label{tab:IMMEoutput} IMME output coefficents (keV) and goodness of fits for lowest $A=20$, $T=2$ quintet.}
\begin{ruledtabular}
\begin{tabular}{c c c c c}
coefficient & quadratic & cubic & quartic only & quartic \\
\hline
$a$               &  9691.1(14)   & 9689.7(17)    & 9690.9(28)   & 9690.9(28)    \\
$b$               &  $-3420.6(5)$ & $-3423.4(20)$ & $-3420.6(5)$ & $-3423.7(21)$ \\
$c$               &  236.5(5)     & 236.8(5)      & 236.9(38)    & 234.4(41)     \\
$d$               &               & 0.8(5)        &              & 0.8(6)        \\
$e$               &               &               &  $-0.1(8)$   & 0.5(9)        \\
\hline
$\chi^2$/$\nu$    &  2.4/2        & 0.28/1        & 2.4/1        &               \\
\end{tabular}
\end{ruledtabular}
\end{table}

Combined with the recently determined mass of $^{20}$Mg from Ref. \cite{ga14prl}, our new value for the isobaric-analog state mass also yields a $Q_{\textrm{EC}}$ value for the superallowed $0^+ \rightarrow 0^+$ transition of $4128.7 \pm 2.2$ keV. This value is sufficiently precise to enable competitive searches for scalar current contributions to nuclear $\beta$ decay using the kinematic broadening of the $^{20}$Mg $\beta$-delayed proton peaks \cite{ad99prl,me13nim}. It can also be used in a precise determination of the $ft$ value for this decay, which would provide a test of the isospin-symmetry breaking calculations used to extract the Cabibbo-Kobayashi-Maskawa (CKM) matrix element $V_{ud}$ from the superallowed decays of $T=1$ nuclides \cite{bh08prc,ha15prc}. More precise values for the half-life of $^{20}$Mg and its superallowed branching are still needed in order to determine a sufficiently precise $ft$ value.

Together with the $A=32$ case \cite{bh08prc}, the present work establishes $\beta$-delayed $\gamma$ decay as a viable method to measure precise and accurate excitation energies for the $T_z = -1$ members of $T=2$ quintets, despite the fact that these states are typically unbound to proton emission by several MeV. For example, we anticipate that this method could be applied to the decays of $^{24}$Si, $^{28}$S, $^{36}$Ca, and so on, given sufficient rare-isotope production and a sufficiently sensitive $\gamma$-ray spectrometer.

In conclusion, recent results indicated that the IMME unexpectedly breaks down in the $A = 20$, $T = 2$ quintet \cite{ga14prl}. Using the $\beta$-delayed $\gamma$ decay of $^{20}$Mg, we measured the excitation energy of the lowest $T = 2$ state of $^{20}$Na. Our value differs by 27 keV from the recommended value and is a factor of 28 more precise. When our new value is adopted in a test of the IMME using the $A = 20$, $T = 2$ quintet, we find that the IMME is revalidated. Therefore, exotic nuclear structure is not currently needed to describe the data on this quintet.

We gratefully acknowledge the NSCL staff for technical assistance and for providing the $^{20}$Mg beam. This work was supported by the National Science Foundation (USA) under Grants No. PHY-1102511, No. PHY-1419765, and No. PHY-1404442.

\bibliographystyle{plainnat}

\end{document}